\RequirePackage{arydshln}
\documentclass[aps,twocolumn,nofootinbib,superscriptaddress,preprintnumbers,pra,10pt,floatfix]{revtex4-2}

\usepackage[utf8]{inputenc}

\usepackage{float}
\usepackage{amsmath,amssymb}
\usepackage{dsfont}
\usepackage{hyperref}
\usepackage{graphicx}
\usepackage{enumitem}
\usepackage{mathtools}
\usepackage{bbold}
\usepackage{multirow}
\usepackage{ytableau}
\usepackage{youngtab}
\usepackage{upgreek}
\usepackage{braket}
\usepackage{soul}
\usepackage{cancel}
\usepackage{xcolor}
\usepackage[normalem]{ulem}
\usepackage[export]{adjustbox}
\usepackage{tikz}
\usepackage{xspace}
\usetikzlibrary{calc,tikzmark,fit,shapes.geometric,matrix,decorations.markings,arrows.meta,decorations.pathmorphing,patterns,positioning,snakes,patterns.meta,calligraphy,decorations.pathreplacing,overlay-beamer-styles}

\usepackage{fontawesome}

\usepackage{bbm}
\usepackage{mathrsfs}
\usepackage{bm}

\usepackage{slashed}

\DeclareUnicodeCharacter{202F}{\ } 
\DeclareUnicodeCharacter{2212}{-} 

\tikzset
  {midarrow/.style={decoration={markings,mark=at position 0.5 with
     {\arrow[thin,xshift=2pt]{Triangle[length=4pt,#1]}}},postaction={decorate}}
  }

\tikzset{
proton/.style = {circle, draw=black, thin, fill=black!20!white, minimum size=#1,
              inner sep=0pt, outer sep=0pt},
proton/.default = 6pt 
}

\tikzset{
blob/.style = {circle, draw=black, thin, preaction={fill, black!20!white}, pattern=north east lines, minimum size=#1,
              inner sep=0pt, outer sep=0pt},
blob/.default = 6pt 
}

\tikzset{
wc/.style = {circle, fill, minimum size=#1,
              inner sep=0pt, outer sep=0pt},
wc/.default = 6pt 
}

\tikzset{vector/.style={decorate, draw=black, decoration={snake=coil,pre length=0, post length = 0}}}

\usepackage{colortbl} 
\definecolor{Gray}{gray}{0.95}
\definecolor{RGray}{gray}{0.290}
\definecolor{CGray}{gray}{0.92}

\newcommand{\belowEW}{$^{(\dagger)}$}
\newcommand{\EE}[1]{\ensuremath{{}\times10^{#1}}}

\usepackage{arydshln}

\definecolor{blueLinks}{RGB}{0,84,159}
\newcommand{\myColor}[0]{blueLinks}
\hypersetup{
	colorlinks, 
	bookmarksopen, 
	bookmarksnumbered,
	pdftoolbar=false, 
	pdfmenubar=false, 
	pdffitwindow=false, 
	pdfstartview={FitH},	 
	citecolor=\myColor, 
	linkcolor=\myColor,	
	urlcolor=\myColor, 
	pdftitle={Probing Lepton Flavor Violation at FCC-ee}, 
	pdfsubject={LFV at FCCee}, 
	pdfnewwindow=true, 
	linktoc=all
}

\makeatletter
\@ifclasslater{revtex4-2}{2022/06/05}{}{%
  \def\@fnsymbol#1{%
    \ensuremath{%
      \ifcase#1\or
        *\or
        \dagger\or
        \ddagger\or
        \mathsection\or
        \mathparagraph\or
        **\or
        \dagger\dagger\or
        \ddagger\ddagger\or
        \mathsection\mathsection\or
        \mathparagraph\mathparagraph\or
        ***\or
        \dagger\dagger\dagger\or
        \ddagger\ddagger\ddagger\or
        \mathsection\mathsection\mathsection\or
        \mathparagraph\mathparagraph\mathparagraph
      \else
        \@ctrerr
      \fi
    }%
  }%
}
\makeatother

\makeatletter
\g@addto@macro\bfseries{\boldmath}
\makeatother

\makeatletter
\renewcommand\paragraph{\@startsection{paragraph}{4}{\z@}%
                                    {3.25ex \@plus1ex \@minus.2ex}%
                                    {-1em}%
                                    {\normalfont\normalsize\bfseries}}
\makeatother

\makeatletter

\makeatletter
\def\@hangfrom@section#1#2#3{\@hangfrom{#1#2}#3}
\def\@hangfroms@section#1#2{#1#2}

\renewcommand{\section}{%
  \@startsection{section}{1}{\z@}%
    {3.5ex \@plus 1ex \@minus .2ex}%
    {2.3ex \@plus .2ex}%
    {\normalfont\centering\fontsize{10.5pt}{12.5pt}\selectfont\bfseries}%
}
\makeatother

\allowdisplaybreaks
\begin{document}

\begin{flushright}
{\small
\preprint{DESY-26-134}
\preprint{P3H-26-084}
\preprint{KA-TP-27-2026}
}
\end{flushright}

\title{Probing Lepton Flavor Violation at FCC-ee}
\author{L.~Allwicher}
\email{lukas.allwicher@desy.de}
\affiliation{Deutsches Elektronen-Synchrotron DESY, Notkestr. 85, 22607 Hamburg, Germany}
\author{R.~Bartocci}
\email{riccardo.bartocci@kit.edu}
\affiliation{Institute for Theoretical Physics, Karlsruhe Institute of Technology, Wolfgang-Gaede-Str. 1, 76131 Karlsruhe,
Germany}
\author{M.~Cepeda}
\email{maria.cepeda@ciemat.es}
\affiliation{Centro de Investigaciones Energ\'eticas, Medioambientales y Tecnol\'ogicas (CIEMAT), Avenida Complutense 40, 28040 Madrid, Spain}
\author{C.~Cornella}
\email{claudia.cornella@pd.infn.it}
\affiliation{Istituto Nazionale di Fisica Nucleare (INFN), Sezione di Padova, Via F.~Marzolo 8, I–35131 Padova, Italy}
\author{L.P.S.~Leal}
\email{luighi.leal@ijclab.in2p3.fr}
\affiliation{IJCLab, P\^ole Th\'eorie (Bat.~210), CNRS/IN2P3 et Universit\'e Paris-Saclay, 91405 Orsay, France}
\affiliation{Instituto de Física, Universidade de São Paulo, São Paulo – São Paulo 05580-090, Brazil}
\author{A.~Lusiani}
\email{alberto.lusiani@sns.it}
\affiliation{INFN, Sezione di Pisa, Pisa, Italy}
\author{C.~Martin~Perez}
\email{cristina.martin@ciemat.es}
\affiliation{Centro de Investigaciones Energ\'eticas, Medioambientales y Tecnol\'ogicas (CIEMAT), Avenida Complutense 40, 28040 Madrid, Spain}
\author{P.~Paradisi}
\email{paride.paradisi@pd.infn.it}
\affiliation{Dipartimento di Fisica e Astronomia ``Galileo Galilei",
Università degli Studi di Padova, Via F. Marzolo 8, 35131 Padova, Italy}
\affiliation{Istituto Nazionale di Fisica Nucleare (INFN), Sezione di Padova, Via F.~Marzolo 8, I–35131 Padova, Italy}
\author{E.~Scantamburlo}
\email{enrico.scantamburlo.3@phd.unipd.it}
\affiliation{Dipartimento di Fisica e Astronomia ``Galileo Galilei",
Università degli Studi di Padova, Via F. Marzolo 8, 35131 Padova, Italy}
\affiliation{Istituto Nazionale di Fisica Nucleare (INFN), Sezione di Padova, Via F.~Marzolo 8, I–35131 Padova, Italy}
\author{O.~Sumensari}
\email{olcyr.sumensari@ijclab.in2p3.fr}
\affiliation{IJCLab, P\^ole Th\'eorie (Bat.~210), CNRS/IN2P3 et Universit\'e Paris-Saclay, 91405 Orsay, France}
%

\begin{abstract}

We investigate the prospects for probing Lepton Flavor Violation (LFV) at FCC-ee within the Standard Model Effective Field Theory (SMEFT) framework. The approximately $2 \times 10^{11}$ $\tau$-lepton pairs expected at the $Z$-pole run of FCC-ee could improve current limits on the branching fractions of $\tau\to3\ell$ and $\tau\to\ell\gamma$ ($\ell=e,\mu$) by factors ranging from 30 to 800, depending on the specific channel. 
We compare the resulting new-physics reach with that obtained from searches for $Z \to \tau \ell$ and $h \to \tau \ell$, which have so far received greater attention in studies of LFV at future $e^+e^-$ colliders. 
We find that $\tau$ decays can surpass both the current and the projected sensitivity of $Z\to\tau\ell$ searches to the same LFV interactions, while probing a broader range of SMEFT operators. 
We further show that, for LFV Yukawa-like interactions, direct searches for $h\to\tau\ell$ provide the strongest sensitivity. By contrast, for semileptonic operators that radiatively induce LFV Higgs decays, this hierarchy can be reversed, with $\tau$ decays providing the leading probe.

\end{abstract}

\maketitle

\allowdisplaybreaks

\newcommand{\cmd}[1]{{\texttt{#1}}\xspace}
\newcommand{\dd}{\mathop{}\!\mathrm{d}}
\newcommand{\braces}[1]{\left\lbrace #1 \right\rbrace}
\newcommand{\brackets}[1]{\left( #1 \right)}
\newcommand{\squarebrackets}[1]{\left[ #1 \right]} 
\newcommand{\angles}[1]{\left\langle #1\right\rangle}
\newcommand{\abs}[1]{\left\lvert #1 \right\rvert}
\newcommand{\norm}[1]{\left\Vert #1 \right\Vert}
\renewcommand{\Im}{\mathop{\mathrm{Im}}}
\renewcommand{\Re}{\mathop{\mathrm{Re}}}
\renewcommand{\L}{\mathcal{L}}
\newcommand{\ord}[1]{\mathcal{O}\!\left( #1 \right)}
\newcommand{\ineqgraphics}[1]{\vcenter{\hbox{\includegraphics[]{#1}}}} 

\newcommand{\SMEFT}[0]{\text{SM\,EFT}\xspace}
\newcommand{\vev}[0]{vacuum expectation value\xspace}
\newcommand{\QFT}[0]{quantum field theory\xspace}
\newcommand{\SU}[1]{\mathrm{SU}(#1)}
\newcommand{\C}[2]{C_{\underset{#2}{#1}}}
\newcommand{\vT}[0]{\overline{v}_T}

\newcommand{\cO}{{\mathcal O}}
\newcommand{\cQ}{{\mathcal Q}}
\newcommand{\cC}{{\mathcal C}}
\newcommand{\cL}{{\mathcal L}}
\newcommand{\cH}{{\mathcal H}}
\newcommand{\cI}{{\mathcal I}}
\newcommand{\cA}{{\mathcal A}}
\newcommand{\cB}{{\mathcal B}}
\newcommand{\cF}{{\mathcal F}}
\newcommand{\cP}{{\mathcal P}}
\newcommand{\cS}{{\mathcal S}}
\newcommand{\cT}{{\mathcal T}}
\newcommand{\cU}{{\mathcal U}}
\newcommand{\cN}{{\mathcal N}}
\newcommand{\cM}{{\mathcal M}}
\newcommand{\cZ}{{\mathcal Z}}

\newcommand{\SMrep}[3]{({\bf #1},\,{\bf #2},\,#3)}
\newcommand{\SMrepbar}[3]{({\bf \bar#1},\,{\bf #2},\,#3)}

\definecolor{myGreen}{rgb}{0.0, 0.7, 0.1}


\section{Introduction}

Lepton Flavor Violation (LFV) is forbidden in the Standard Model (SM) as a consequence of the accidental global symmetry $U(1)_e \times U(1)_\mu \times U(1)_\tau$, which implies the conservation of lepton family numbers. The observation of neutrino oscillations, however, demonstrates that this symmetry is violated in nature. Nevertheless, the contributions of light neutrino masses to charged 
LFV processes are suppressed by powers of $m_\nu^2/m_W^2$~\cite{Petcov:1976ff,Lee:1977tib,Marciano:1977wx}, and therefore lie 
far beyond the reach of present and foreseeable experiments. Any observable signal of charged LFV would then provide unambiguous evidence for new physics.

Current experiments probe charged LFV in both $\mu \to e$ and $\tau \to e,\,\mu$ transitions~\cite{Calibbi:2017uvl}. 
In flavor-anarchic new-physics scenarios, the most stringent constraints arise from muon processes such as $\mu \to e \gamma$~\cite{MEGII:2025gzr}, $\mu \to 3e$~\cite{SINDRUM:1987nra}, and coherent $\mu\to e$ conversion in nuclei~\cite{SINDRUMII:2006dvw}. With current data, these observables already test extremely high effective scales, up to $\mathcal{O}(10^3~\text{TeV})$~\cite{EliasMiro:2021jgu,Davidson:2022jai,Fernandez-Martinez:2024bxg,Plakias:2023esq}. Future facilities, including COMET~\cite{COMET:2018auw} and Mu2e~\cite{Mu2e:2014fns}, are expected to further improve this sensitivity by up to two orders of magnitude. 
The interpretation of these bounds, however, depends on the flavor structure of the underlying theory.
In particular, scenarios featuring hierarchical couplings that predominantly involve third-generation leptons can substantially weaken the constraints from $\mu \to e$ transitions. This well-motivated assumption makes LFV processes involving $\tau$ leptons complementary and, in some cases, leading probes of the underlying flavor structure of new physics~\cite{Barbieri:1995uv,Barbieri:1996ww,Barbieri:1997tu, Kagan:2009bn,Barbieri:2012uh,Panico:2015jxa,Panico:2016ull,Barbieri:2012tu,Glioti:2024hye,Bordone:2017bld,Fuentes-Martin:2022xnb, Faroughy:2020ina,Davighi:2023iks}.

At present, the most stringent limits on $\tau$-lepton LFV decays are provided by $B$-factories~\cite{ParticleDataGroup:2024cfk}. 
In recent years, LHC experiments have also made significant progress, particularly in the $\tau \to 3 \mu$ channel~\cite{CMS:2023iqy,LHCb:2026eod}, reaching sensitivities competitive with those reported by Belle~II~\cite{Belle-II:2024sce}.
Looking ahead, Belle~II has the potential to improve the sensitivity to LFV $\tau$-lepton decay modes by about one order of magnitude~\cite{Belle-II:2018jsg}. Among the proposed future colliders, a high-luminosity $e^+e^-$ collider operating at the $Z$ pole, such as FCC-ee~\cite{FCC:2025lpp} or CEPC~\cite{CEPCStudyGroup:2018ghi}, offers particularly strong prospects for these searches, given the clean experimental environment and the large number of $\tau$ pairs produced. Indeed, for a sample of about $6\times 10^{12}$ $Z$ bosons, FCC-ee would produce about $2\times 10^{11}$ $\tau$ pairs~\cite{FCC:2025lpp}, enabling many opportunities to go beyond the expected physics reach of Belle II in the $\tau$ sector~\cite{deBlas:2025gyz}. FCC-ee is also expected to operate at $\sqrt{s}=240~\mathrm{GeV}$,  where $\mathcal{O}(10^6)$ Higgs bosons are expected to be produced in association with a $Z$-boson, $e^+e^-\to Zh$, allowing for searches for LFV Higgs decays. In this channel, the clean experimental environment allows the Higgs boson to be reconstructed through the recoil mass against an on-shell $Z$ boson. Preliminary projections indicate that FCC-ee could surpass the HL-LHC sensitivity to LFV Higgs decays~\cite{Qin:2017aju}.

Previous studies of LFV processes at high-luminosity $e^+e^-$ facilities have mainly focused on $Z$-boson~\cite{Calibbi:2021pyh,Abada:2014cca,Abada:2015zea} and Higgs-boson decays~\cite{Sriling:2026vmx}. Within the SM Effective Field Theory (SMEFT) framework, it has been shown that these channels can probe specific LFV interactions with a reach comparable to that of $\tau$-lepton decays at Belle~II~\cite{Calibbi:2021pyh}. 
In this work, we show that this comparison changes once $\tau$-lepton decays accessible at the same Tera-$Z$ facility are taken into account. More specifically, $\tau \to \ell \gamma$ and $\tau \to 3 \ell$ ($\ell = e,\mu$) can provide stronger probes of several SMEFT interactions that induce $Z\to\tau \ell$, while at the same time testing a broader set of operators.  

In this letter, we analyze the reach of LFV searches at FCC-ee within the SMEFT framework. We first compare searches for $Z \to \tau \ell$ with LFV $\tau$-lepton decays accessible during the same Tera-$Z$ run. We then study LFV Higgs interactions, comparing direct searches for $h\to\tau\ell$ at $\sqrt{s}=240~\mathrm{GeV}$ with indirect constraints from LFV $\tau$ decays at the $Z$ pole. 

The remainder of this letter is organized as follows. In Sec.~\ref{sec:exp}, we collect the current bounds and future projections for the LFV processes relevant to our analysis, revisiting and refining these estimates when necessary. In Sec.~\ref{sec:eft}, we introduce our notation and formulate the SMEFT description of charged LFV interactions. 
In Sec.~\ref{sec:pheno}, we compare the reach for LFV processes at a future $e^+ e^-$ collider, focusing on the interplay of $\tau$ decays with LFV $Z$ and Higgs decays. We summarize our main findings in Sec.~\ref{sec:outlook}.

\section{Experimental prospects for LFV}
\label{sec:exp}

In this section, we summarize the experimental inputs used in the SMEFT analysis below.
We collect the current bounds and the best projections at existing and planned facilities, namely Belle II for LFV $\tau$-lepton decays and the HL-LHC for $Z$- and Higgs-boson decays. These limits are then compared with the expected reach of FCC-ee. All projections, along with current bounds, are summarized in Table~\ref{tab:exp}.

\paragraph*{Projections at the HL-LHC and Belle II.}
For Belle~II, projections correspond to an integrated luminosity of $50~\mathrm{ab}^{-1}$, while HL-LHC projections refer to $3~\mathrm{ab}^{-1}$.  
For $\tau$-lepton decays we use the projections quoted in Ref.~\cite{Belle-II:2022cgf}. 
For Higgs-boson decays we use the projections of Ref.~\cite{ATLAS:2022alf}. 
For $Z$ decays we extrapolate existing ATLAS bounds~\cite{ATLAS:2021bdj} to the HL-LHC luminosity, obtaining
\begin{align}
\begin{aligned}
\mathcal{B}(Z \to e \tau )_{\text{HL-LHC}} < 1.2 \times 10^{-6}\,,\\ 
\mathcal{B} (Z \to \mu \tau )_{\text{HL-LHC}} < 1.5 \times 10^{-6}\,, 
\end{aligned}
\end{align}
at $95\%$~C.L.~\footnote{Throughout this work, $\mathcal{B}(Z \to \tau \ell)= \mathcal{B}(Z \to \tau^- \ell^+)+ \mathcal{B}(Z \to \tau^+ \ell^-)$. The same holds for $h \to \tau \ell$.}  These projections assume statistical scaling with the integrated luminosity, motivated by the fact that current limits are still largely statistically limited. They should therefore be regarded as optimistic estimates, as the increasing impact of systematic uncertainties at higher luminosities is neglected.

\paragraph*{Projections at FCC-ee.}
For LFV $\tau$ and $Z$ decays, we consider a Tera-$Z$ run with $6\times10^{12}$ produced $Z$ bosons, corresponding to approximately $2\times10^{11}$ $\tau$-lepton pairs. For LFV Higgs decays, we assume a $\sqrt{s}=240~\mathrm{GeV}$ run with $1.2\times10^6$ produced Higgs bosons~\cite{FCC:2025lpp}. Several of the FCC-ee projections rely on extrapolations or on studies based on limited simulated samples. We briefly summarize their derivation below.

\begin{itemize}[leftmargin=*]
\item  \boldsymbol{$\tau \to \ell \gamma$}. 
Dedicated FCC-ee studies are available only for $\tau \to \mu \gamma$~\cite{Dam:2018rfz, lusiani_2024_57pxj-6xd43-article}. For this channel, the FCC-ee sensitivity is expected to be limited by Poisson fluctuations in the background yield. The main background was estimated using Monte Carlo samples of $\tau \to \mu \bar{\nu}\nu$ decays in $Z \to \tau^+\tau^-(\gamma)$ events, with the generated events smeared to account for the expected detector resolution. By extrapolating the number of background events passing the signal selection from a sample corresponding to 2\% of the full FCC-ee $Z$-peak statistics, a two-standard-deviation sensitivity was estimated~\cite{Dam:2018rfz}. In 2024, this estimate was rescaled to the updated FCC-ee $Z$-peak sample size and used to calculate an expected 90\% C.L.~upper limit of~\cite{lusiani_2024_57pxj-6xd43-article}
\begin{align}
\mathcal{B}(\tau \to \mu \gamma)_{\text{FCC-ee}} < 1.2 \times 10^{-9}\,.
\end{align}

In the absence of a dedicated study of the electron mode, we assume the same projected sensitivity for $\tau\to
e\gamma$ as an optimistic benchmark.

\item  \boldsymbol{$\tau \to 3 \ell$}. The prospects for $\tau \to 3 \mu$ and $\tau \to 3 e$ at FCC-ee have recently been estimated using fast simulations\footnote{Monte Carlo event generation followed by a parametric description of the detector response, digitization, and reconstruction.} of signal and background samples \cite{Kazanecki_2026_phipsi26}. Events were selected using dedicated criteria, including the requirement that the invariant mass and total energy of the three reconstructed leptons be consistent with the $\tau$ mass and its expected energy. 
The simulated samples corresponded to only about half per mille of the full FCC-ee design sample, although event-weighting techniques increase the effective statistical power in some cases. Bayesian inference with Jeffreys priors was used to account for non-zero expected backgrounds even when no simulated events survive the selection. The resulting expected $90\%$ C.L. upper limits are~\cite{Kazanecki_2026_phipsi26}
\begin{align}
    \begin{alignedat}{2}
    &\mathcal{B}(\tau^- \to \mu^-\mu^+\mu^-)_{\text{FCC-ee}} &{}< 2.4\EE{-11}\,,
    \\
    &\mathcal{B}(\tau^- \to e^-e^+e^-)_{\text{FCC-ee}} &{}< 5.1\EE{-10}\,.    
    \end{alignedat}
\end{align}
The first limit updates an earlier estimate of
$2.0\EE{-11}$~\cite{lusiani_2024_57pxj-6xd43-article}, obtained by extrapolating
the recent Belle-II search~\cite{Belle-II:2024sce}. 

\begin{table}[!t]
    \renewcommand{\arraystretch}{2.3}
    \centering
        \centering
 \begin{tabular}{c|cc|cc|cc}

    Process &   Current   & &  Future   &  &   FCC-ee   & 
    \\\hline\hline
    $\tau\to e \gamma$    &  $3.3\times 10^{-8}$ & \cite{BaBar:2009hkt} &     $9.1 \times 10^{-9}$ & \cite{Belle-II:2022cgf} & $1.2\times 10^{-9}$ & $^{(\ast)}$ \\
    $\tau\to \mu \gamma$   & $4.2\times 10^{-8}$ & \cite{Belle:2021ysv} & 
    $6.9\times 10^{-9}$& \cite{Belle-II:2022cgf} & $1.2\times 10^{-9}$& \cite{lusiani_2024_57pxj-6xd43-article}\\ \hline
    $\tau\to 3e$    & $2.7\times 10^{-8}$ & \cite{Hayasaka:2010np} & $4.7\times 10^{-10}$ & \cite{Belle-II:2022cgf}& $5.1\times 10^{-10}$ & \cite{Kazanecki_2026_phipsi26} \\
    $\tau\to 3\mu$  & $1.9\times 10^{-8}$ & \cite{Belle-II:2024sce} & $4.5\times 10^{-10}$ & \cite{Belle-II:2022cgf} & $2.4\times 10^{-11}$  & \cite{Kazanecki_2026_phipsi26}\\    \hline\hline
$Z\to e\tau$    &      $5.0\times 10^{-6}$ & \cite{ATLAS:2021bdj} & {$1.2\times 10^{-6}$}  &  $^{(\ast)}$ & $7.1 \times 10^{-10}$ & $^{(\ast)}$\\
 $Z\to \mu\tau$   &  $6.5\times 10^{-6}$ & \cite{ATLAS:2021bdj} &  {$1.5\times 10^{-6}$}  & $^{(\ast)}$  & $7.1 \times 10^{-10}$ & $^{(\ast)}$\\ \hline\hline
    $h\to e\tau$    & $2.0\times 10^{-3}$ & \cite{ATLAS:2023mvd} & $2.4\times 10^{-4}$ & \cite{ATLAS:2022alf}  
    & $9.2 \times 10^{-5}$   & $^{(\ast)}$ \\ 
    $h\to\mu\tau$  & $1.5\times 10^{-3}$ & \cite{CMS:2021rsq} & $2.4\times 10^{-4}$  & \cite{ATLAS:2022alf}  & $8.1 \times 10^{-5}$ & $^{(\ast)}$ \\    
 \end{tabular}
    \caption{ 
    \small  \sl
    Current upper limits on the branching fractions of LFV $\tau$, $Z$, and Higgs decays are compared to projected sensitivities at Belle II, the HL-LHC, and FCC-ee.    Limits and projected sensitivities for $\tau$-lepton decays are quoted at $90\%$ C.L., whereas those for $Z$- and Higgs-boson decays are quoted at $95\%$ C.L.  Projections labeled with an asterisk have been reassessed by us as explained in the text. } 
\label{tab:exp} 
\end{table}

\item \boldsymbol{$Z \to \tau \ell$}. The FCC-ee reach for $Z \to \tau \ell$ was estimated in Ref.~\cite{Dam:2018rfz}, which studies the $Z\to e\tau$ and $Z\to\mu\tau$ channels using multi-prong hadronic $\tau$-lepton decays and exploits the excellent momentum resolution expected at FCC-ee to suppress the dominant background from $Z\to\ell\ell$, as well as the irreducible one from $Z\to\tau\tau$. It finds sensitivities of order $\mathcal{B}(Z\to\tau\ell) < 1 \times 10^{-9}$ for $3 \times 10^{12}$ visible $Z$ decays. We rescale this projection to a $6\times 10^{12}$ $Z$-boson sample assuming statistical scaling, leading to 
\begin{equation}
    \mathcal{B}(Z\to\tau\ell)_{\text{FCC-ee}} < 7.1 \times 10^{-10}\,.
\end{equation}

\item \boldsymbol{$h \to \tau \ell$}. The prospects for $h\to\tau\ell$ at FCC-ee have been investigated in Refs.~\cite{Qin:2017aju,Sriling:2026vmx}. We reassess these projections using the $\tau$-reconstruction performance obtained in a recent FCC-ee study of $h\to\tau\tau$~\cite{Giappichini:2026vlg}. The resulting sensitivity benefits primarily from two ingredients. The first is the inclusion of hadronically decaying $\tau$ leptons. Previous analyses considered only leptonic $\tau$ decays, accounting for about 35\% of the total $\tau$ branching fraction. Including hadronic decays therefore increases the available signal yield by a factor of $\approx 2.86$ and improves the expected branching-fraction sensitivity by a factor of $\approx 1.69$, while benefiting from the high efficiency and good background rejection of modern $\tau$ reconstruction algorithms. The second is the combination of statistically independent $Z$-reconstruction categories. Previous analyses considered the same LFV Higgs final states, but target hadronic ($Z\to qq$) and leptonic ($Z\to\ell\ell$) $Z$-boson decays individually. Assuming statistical independence, their expected limits can be combined in inverse quadrature. Including both these changes yields the following projected $95\%$ C.L. limits:
\begin{align}
\begin{aligned}
\mathcal{B}(h\to e\tau)_{\text{FCC-ee}} &< 9.2\times 10^{-5}\,, \\
\mathcal{B}(h\to \mu\tau)_{\text{FCC-ee}} &< 8.1\times 10^{-5}\,,
\end{aligned}
\end{align}
which improve over the corresponding HL-LHC projections~\cite{ATLAS:2022alf} by approximately a factor of three.

\end{itemize}
Several comments follow from Table~\ref{tab:exp}. Compared with the projected sensitivities at Belle~II and the HL-LHC, the largest improvement at FCC-ee is expected for LFV $Z$ decays, where bounds could improve by more than three orders of magnitude. Among LFV $\tau$ decays, the largest gain is expected for $\tau\to3\mu$, which can surpass the Belle-II projection by more than one order of magnitude. A moderate but still significant improvement is expected for $\tau \to \ell \gamma$ and for $h\to \tau \ell$ decays: FCC-ee could improve the sensitivity by a factor of  $\approx 6$--$8$ for the former and $\approx 3$ for the latter.  While all these projections await dedicated full-detector simulations, they are nevertheless encouraging and motivate the new-physics study presented below.

\section{LFV in the SMEFT}
\label{sec:eft}

We assume that new physics lies at energy scales well above the electroweak scale, so that its effects at low energy can be systematically described in the SMEFT~\cite{Buchmuller:1985jz}, 
%
\begin{align}   
    \mathcal{L}_{\mathrm{SMEFT}} \supset \sum_a \mathcal{C}_a\,\mathcal{O}_a\,,
\end{align}
%
where $\mathcal{C}_a$ are (dimensionful) Wilson coefficients of the effective operators $\mathcal{O}_a$. Restricting ourselves to LFV interactions, the leading operators appear at dimension six, which we take in the so-called Warsaw basis~\cite{Grzadkowski:2010es}.

For the processes considered in this work, four classes of SMEFT operators are relevant: (i) dipole operators ($\bm{\psi^2 X H}$), (ii) Higgs-current operators ($\bm{\psi^2 H^2 D}$), (iii) Yukawa-like operators ($\bm{\psi^2 H^3}$), and (iv) leptonic and semileptonic four-fermion operators ($\bm{\psi^4}$). The operators retained in our analysis are summarized in Table~\ref{tab:tau-ope-smeft}. In the semileptonic sector, we restrict ourselves to $\mathcal{O}_{ledq}$ and $\mathcal{O}_{l equ}^{(1)}$ with third-generation quark flavor indices, which generate LFV Higgs decays at one loop, as discussed in Sec.~\ref{subsec:LFV-H-tau-decays}.~\footnote{A systematic analysis of semileptonic LFV $\tau$ decays at FCC-ee is beyond the scope of this work. For constraints on semileptonic SMEFT LFV operators from existing facilities, including vector and tensor operators, see Refs.~\cite{Plakias:2023esq,Angelescu:2020uug,Fernandez-Martinez:2024bxg} and references therein.} A given SMEFT operator generally contributes to several of the processes considered here, thereby inducing correlations among different observables.

Here we consider scenarios in which one or two of the operators in Table~\ref{tab:tau-ope-smeft} have non-zero Wilson coefficients at a large scale $\Lambda$. We then derive the constraints from the different LFV decays as follows. The Wilson coefficients are evolved from $\Lambda$ down to the electroweak scale using one-loop renormalization group (RG) equations~\cite{Jenkins:2013wua,Jenkins:2013zja,Alonso:2013hga}. At the electroweak scale, the SMEFT is matched onto the Low-Energy Effective Field Theory (LEFT)~\cite{Jenkins:2017jig}, in which the heavy degrees of freedom ($W$, $Z$, $h$, and the top quark) are integrated out. For $\tau$ decays, the resulting LEFT Wilson coefficients are evolved down to the $\tau$ mass via one-loop RG equations~\cite{Jenkins:2017dyc}. The one-loop RG evolution below and above the electroweak scale is implemented using \texttt{DsixTools}~\cite{Fuentes-Martin:2020zaz}. 

For each observable, we retain the leading phenomenologically relevant contribution in the perturbative expansion. Besides tree-level effects, this can include the one-loop matching \cite{Dekens:2019ept} of the SMEFT to the LEFT and one-loop contributions to matrix elements where relevant, as well as higher-order effects when they provide the leading sensitivity to a given operator. In particular, two-loop and logarithmically enhanced higher-order contributions are relevant for radiative $\tau$ decays induced by the Yukawa-like and semileptonic operators. The expressions for the observables and the contributions included in our analysis are collected in Appendix~\ref{app:observables}.

\begin{table}[t!]
    \renewcommand{\arraystretch}{1.85}
    \centering
    \begin{tabular}{c|c||c|c}
       \multicolumn{2}{c||}{$\bm{\psi^2 H^2 D}$}    & \multicolumn{2}{c}{$\bm{\psi^4}$} \\ \hline
      \multirow{2}{*}{$\mathcal{O}_{Hl}^{(1)}$}  & \multirow{2}{*}{$\big{(}H^\dagger i\overleftrightarrow{D}_\mu H\big{)} \big{(}\bar{l} \gamma^\mu l\big{)}$}  &  $\mathcal{O}_{l l}$ & $\big{(}\bar{l} \gamma^\mu l\big{)}\big{(}\bar{l}\gamma_\mu l\big{)}$\\[0.3em]
      &&  $\mathcal{O}_{l e}$ & $\big{(}\bar{l} \gamma^\mu l\big{)}\big{(}\bar{e}\gamma_\mu e\big{)}$ \\[0.3em]
      $\mathcal{O}_{Hl}^{(3)}$  &  $\big{(}H^\dagger i\overleftrightarrow{D}_\mu^I H\big{)} \big{(}\bar{l} \gamma^\mu \tau^I l\big{)}$ &  $\mathcal{O}_{ee}$ & $\big{(}\bar{e} \gamma^\mu e\big{)}\big{(}\bar{e}\gamma_\mu e\big{)}$\\[0.3em] 
      \multirow{2}{*}{$\mathcal{O}_{He}$}  &  \multirow{2}{*}{$\big{(}H^\dagger i \overleftrightarrow{D}_\mu H\big{)} \big{(}\bar{e} \gamma^\mu e\big{)}$}& $\mathcal O_{l edq}$&$\left(\bar{l} e\right)\left( \bar{d} q\right)$\\[0.3em]
      &&$\mathcal O_{l equ}^{(1)}$&$\left(\bar{l} e\right)i \tau_2 \left( \bar{q} u\right)$\\[0.3em]
      \hline\hline
       \multicolumn{2}{c||}{$\bm{\psi^2 X H}+\mathbf{h.c.}$}    & \multicolumn{2}{c}{$\bm{\psi^2 H^3}+\mathbf{h.c.}$}    \\ \hline
       $\mathcal{O}_{eB}$ &  $\big{(}\bar{l} \sigma^{\mu\nu} e \big{)}\,H B_{\mu\nu}$ &  \multirow{2}{*}{$\mathcal{O}_{eH}$} & \multirow{2}{*}{$\big{(}H^\dagger H\big{)}\big{(}\bar{l} e \big{)}\,H $} \\[0.3em]
       $\mathcal{O}_{eW}$ &  $\big{(}\bar{l} \sigma^{\mu\nu} \tau^I e \big{)}\,H W_{\mu\nu}^I$ &   & \\
    \end{tabular}
    \caption{\small \sl Dimension-six operators considered in our analysis. Flavor indices are omitted, $SU(2)_L$ indices are denoted by capital letters, and $\tau^I$ ($I=1,2,3$) are the Pauli matrices. See Appendix~\ref{app:conventions} for our
notation and conventions.}
    \label{tab:tau-ope-smeft} 
\end{table}

\section{New-physics reach of LFV searches at FCC\lowercase{-ee}}
\label{sec:pheno}

We now compare the new-physics reach of the LFV observables discussed in Sec.~\ref{sec:exp} by mapping their projected sensitivities onto the SMEFT operators listed in Table~\ref{tab:tau-ope-smeft}. 

Starting with $Z$- and Higgs-boson decays, which have received most of the attention in studies of LFV at FCC-ee, we ask whether leptonic $\tau$ decays can provide comparable or stronger constraints on the same underlying interactions. 
To this end, we identify the SMEFT operators contributing to $Z \to \tau \ell$ and to $h \to \tau \ell$ at a given perturbative order, and then determine their contributions to $ \tau \to 3 \ell$ and $ \tau \to \ell \gamma$ decays, comparing the resulting constraints with those from the corresponding direct $Z$ and Higgs searches. 

We focus on the $\tau\leftrightarrow \mu$ sector. The corresponding results for $\tau\leftrightarrow e$ transitions lead to qualitatively very similar conclusions, with minor differences arising from the experimental sensitivities and from the dependence of the observables on the light lepton masses, cf.~Appendix~\ref{app:observables}.

\subsection{LFV $Z$ and $\tau$ decays}
\label{subsec:LFV-Z-tau-decays}

\begin{figure*}[!t]
\centering
    \includegraphics[width=0.45\textwidth]{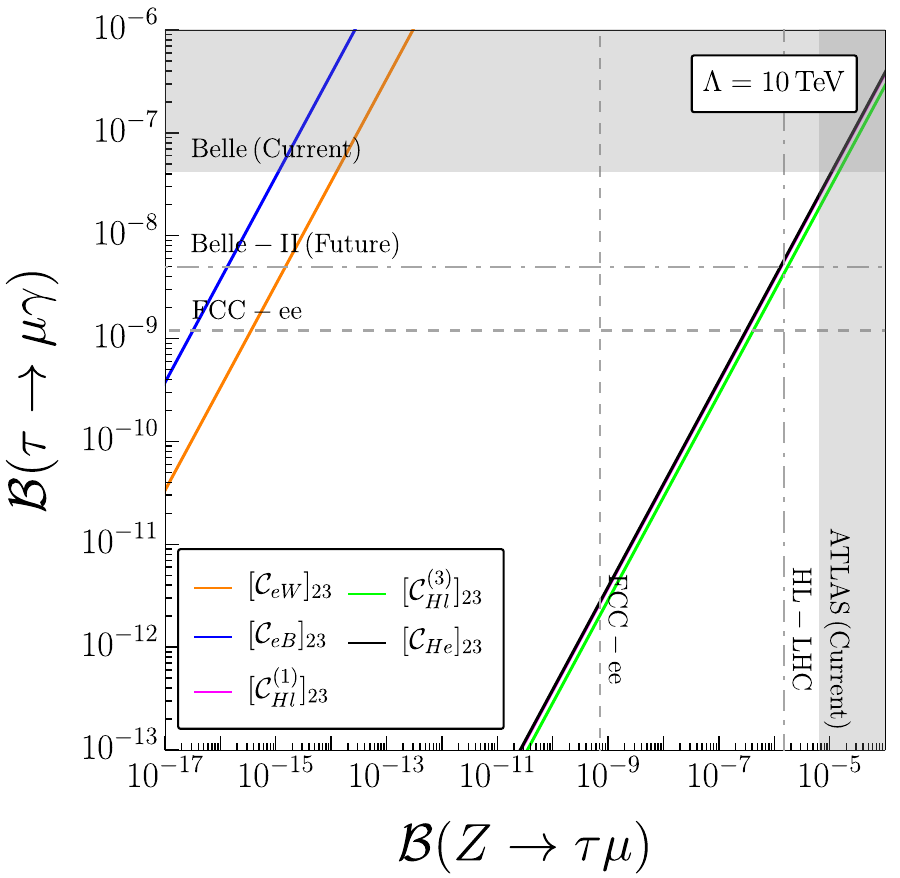}
    ~\quad~ \includegraphics[width=0.45\textwidth]{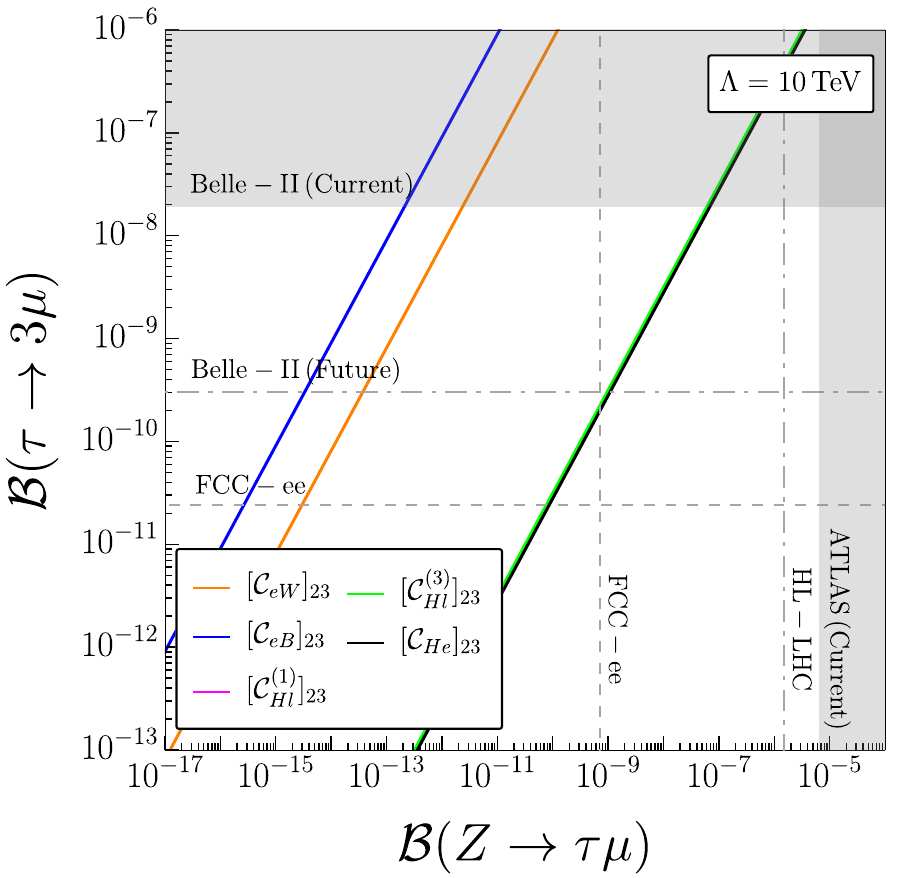}
    \caption{\small \sl Correlations between $\mathcal{B}(Z \to \tau \mu)$ and~$\mathcal{B}(\tau \to \mu\gamma)$ (left panel) and between $\mathcal{B}(Z \to \tau \mu)$ and~$\mathcal{B}(\tau \to 3\mu)$ (right panel), obtained by varying the Wilson coefficients $\mathcal{C}_{eB}$ (blue), $\mathcal{C}_{eW}$ (orange), $\mathcal{C}_{He}$ (black), $\smash{\mathcal{C}_{Hl}^{(1)}}$ (magenta) and $\smash{\mathcal{C}_{Hl}^{(3)}}$ (green), with lepton flavor indices $23$, namely those relevant for $\tau\to \mu$ transitions. The current constraints (gray regions) and projected sensitivities at Belle~II and FCC-ee (dashed lines), as per Table~\ref{tab:exp}, are shown for comparison. The scale $\Lambda$ from which running effects are taken into account is fixed to be $10$~TeV.}
    \label{fig:correlation-plots-Z}
\end{figure*}
\begin{figure}[h]
    \centering    
    \begin{tikzpicture}[thick,>=stealth]
        \draw[vector] (-1.5,0) node[left] {$Z$} -- (0,0);
        \draw[midarrow] (0,0) -- +(45:1.2) node[right] {$\tau$};
        \draw[midarrow] (0,0) +(-45:1.2) node[right] {$\ell$} -- (0,0);
        \draw[dotted] (0,0) -- +(135:1.2) node[left] {$v$};
        \draw[dotted] (0,0) -- +(225:1.2) node[left] {$v$};
        \fill[] (0,0) circle (2pt);
    \end{tikzpicture}
    \begin{tikzpicture}[thick,>=stealth]
        \draw[vector] (0,0) -- (1.5,0);
        \draw[midarrow] (0,0) -- +(225:1.2) node[left] {$\ell$};
        \draw[midarrow] (0,0) +(135:1.2) node[left] {$\tau$} -- (0,0);
        \draw[dotted] (0,0) -- +(45:1.2) node[right] {$v$};
        \draw[dotted] (0,0) -- +(-45:1.2) node[right] {$v$};
        \fill[] (0,0) circle (2pt);
        \node[above] at (1,0) {$Z$};
        \draw[midarrow] (1.5,0) -- +(45:1.2) node[right] {$\ell$};
        \draw[midarrow] (1.5,0) +(-45:1.2) node[right] {$\ell$} -- (1.5,0);
    \end{tikzpicture}
    \caption{\small  \sl
    Tree-level contributions of Higgs-current operators ($\bm{\psi^2H^2D}$) to $Z\to\tau\ell$ and $\tau\to3\ell$ through an LFV $Z$-boson coupling. Similar diagrams can be drawn for dipole operators ($\bm{\psi^2 X H}$).}
    \label{fig:lfv-Z}
\end{figure}
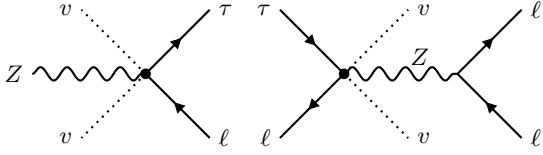

At tree level, the $Z \to \tau \mu$ decay receives contributions from the Higgs-current operators $\mathcal{O}_{Hl}^{(1)}$, $\mathcal{O}_{Hl}^{(3)}$ and $\mathcal{O}_{He}$, as well as from the dipole operators $\mathcal{O}_{eB}$ and $\mathcal{O}_{eW}$. 
In both cases, the resulting $Z\tau\mu$ interaction mediates $\tau\to 3\mu$ at tree level through the exchange of a virtual $Z$ boson and the insertion of a flavor-conserving SM coupling of the $Z$ to leptons. This mechanism is illustrated for the Higgs-current operators in Fig.~\ref{fig:lfv-Z}.
In addition, $\mathcal{O}_{eB}$ and $\mathcal{O}_{eW}$ generate an electromagnetic dipole, which induces $\tau\to\mu\gamma$ and $\tau\to 3\mu$, the latter through off-shell photon exchange. 
The contributions from the Higgs-current operators to $\tau\to\mu\gamma$ instead occur via two different one-loop effects: a finite matching contribution to the electromagnetic dipole, and a one-loop matrix element of the four-lepton operators generated by tree-level $Z$ exchange. Further details are given in Appendix~\ref{app:observables}.

We first illustrate these correlations in Fig.~\ref{fig:correlation-plots-Z}. For each of the SMEFT operators contributing to $Z\to\tau\mu$ at tree level, we switch on the corresponding Wilson coefficient at $\Lambda=10$~TeV and vary its magnitude. The figure shows the resulting correlations between $\mathcal{B}(Z\to\tau\mu)$ and $\mathcal{B}(\tau\to\mu\gamma)$ (left panel), and between $\mathcal{B}(Z\to\tau\mu)$ and $\mathcal{B}(\tau\to3\mu)$ (right panel). The RG evolution from $\Lambda=10~\mathrm{TeV}$ down to the characteristic scale of the processes considered is included. Current constraints and the projected sensitivities at Belle~II and FCC-ee are shown for comparison.

From the plots, it is clear that, already with present $B$-factory data, $\tau$-lepton decays provide stronger constraints than searches for $Z \to \tau \mu$ at the LHC. Although at FCC-ee the sensitivity to $Z \to \tau \mu$ is expected to improve by almost four orders of magnitude with respect to the current LHC bound, significantly more than LFV $\tau$ decays, projections indicate that $\tau$-lepton decays will continue to be the most sensitive probes of LFV couplings of the $Z$ boson.

This conclusion is also reflected in the individual constraints presented in Fig.~\ref{fig:1D-constraints}.
For each single coefficient scenario, the reach in the new-physics scale is determined by setting one coefficient at a time to $\mathcal{C}_a(\Lambda)=1/\Lambda^2$, where $\Lambda$ is the scale from which RG evolution is performed. For Higgs-current operators, $\tau\to 3\mu$ probes scales of about $40$~TeV at FCC-ee, compared with about $25$~TeV from direct $Z\to\tau\mu$ searches. For dipole operators, the hierarchy is much stronger: $\tau\to\mu\gamma$ reaches scales of order $10^3~\mathrm{TeV}$, while the direct $Z$ decay probes only $\mathcal{O}(20)$~TeV. The weaker sensitivity of $\tau\to\mu\gamma$ to Higgs-current operators reflects the fact that their contribution to the process arises only at one loop. The numerical values of the scale reaches shown in Fig.~\ref{fig:1D-constraints} are collected in Table~\ref{tab:numerical_scale_bounds} in  Appendix~\ref{app:2D-SMEFT}. 

These conclusions do not change for operators that contribute to $Z \to \tau \mu$ radiatively. In the most favorable case, the same loop-induced $Z \tau \mu$ vertex contributes to both $Z \to \tau \mu$ and $\tau \to 3\mu$, the latter through off-shell $Z$ exchange (cf.~Fig.~\ref{fig:lfv-Z}) and without any additional loop suppression. The fixed relation between the two processes, together with the stronger sensitivity to $\tau\to3\mu$, still leads to the latter providing the tighter constraint.  

\begin{figure*}[t!]
\centering
    \includegraphics[width=0.45\textwidth]{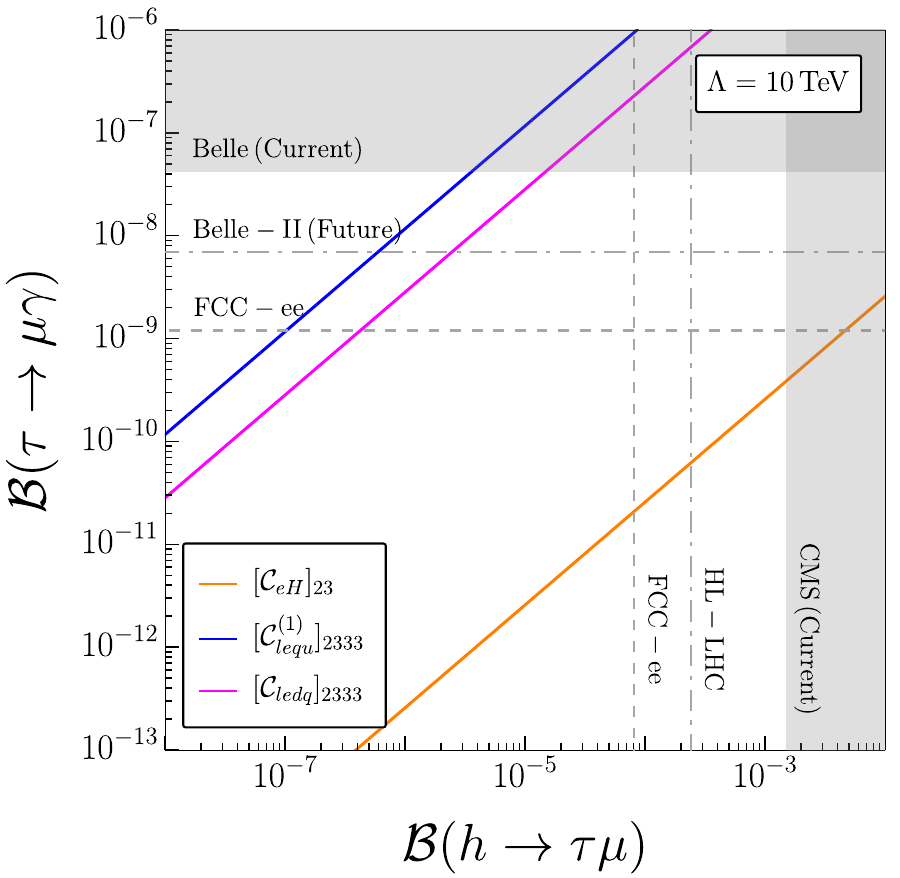}
    ~\quad~\includegraphics[width=0.45\textwidth]{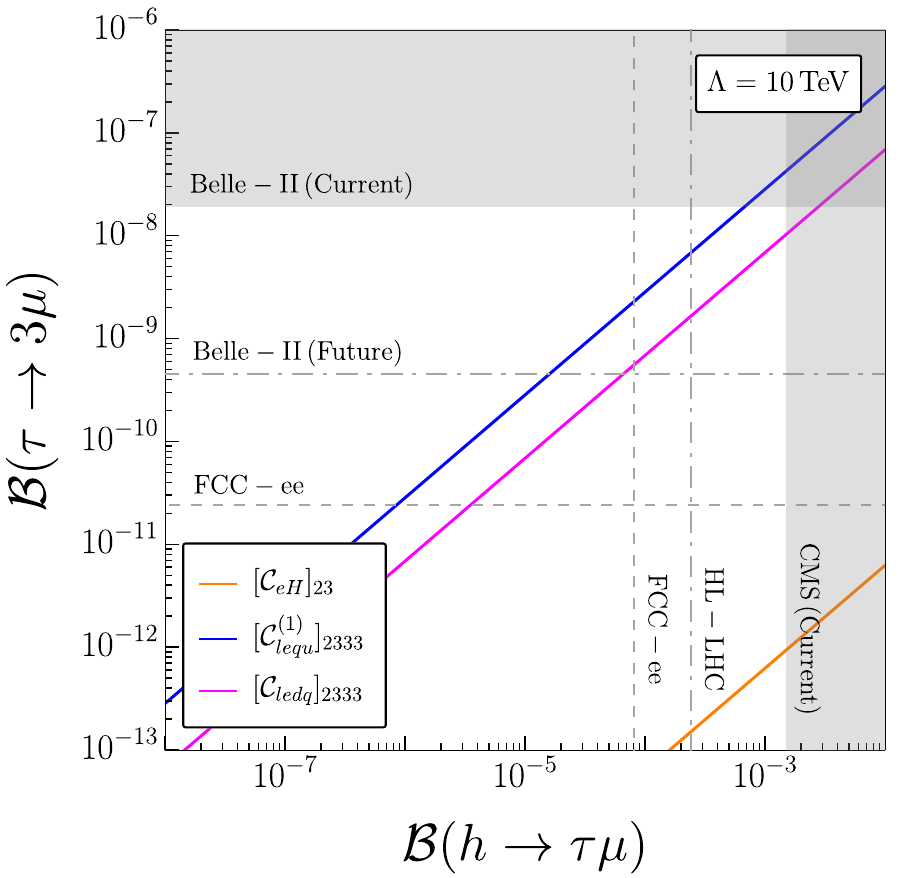}
    \caption{\small \sl Correlations between $\mathcal{B}(h \to \tau \mu)$ vs.~$\mathcal{B}(\tau \to \mu\gamma)$ (left panel) and $\mathcal{B}(h \to \tau \mu)$ vs.~$\mathcal{B}(\tau \to 3\mu)$ (right panel), obtained by varying the Wilson coefficients $\mathcal{C}_{eH}$ (orange), $\mathcal{C}_{l e qu}^{(1)}$ (blue) and $\mathcal{C}_{ledq}$ (magenta), with lepton flavor indices $23$ and quark indices $33$. The expected FCC-ee sensitivity to the corresponding channels is depicted by dashed lines, which are compared to the current constraints (gray regions), as well as to the future sensitivity of Belle II, as per Table~\ref{tab:exp}.}
    \label{fig:correlation-plots-H}
\end{figure*}

For other operators, the hierarchy can be even less favorable to $Z \to \tau \mu$. This occurs, for example, when the effective $Z\tau\mu$ vertex is generated at loop level by a four-fermion operator of the type $(\bar\mu\gamma_\alpha \tau)(\bar\psi \gamma^\alpha \psi)$ by closing the $\psi$ loop with an external $Z$ line. If the fermions in the loop are light enough to be produced in $\tau$ decays, the same four-fermion interaction can mediate an LFV $\tau$-lepton decay directly at tree level, and the latter would provide the strongest constraint. 
This is the case for the four-lepton operators $\mathcal{O}_{ll}$, $\mathcal{O}_{ee}$, and $\mathcal{O}_{le}$.  They contribute to $\tau\to 3\mu$ already at tree level and affect $Z \to \tau \mu$ only at loop level, predominantly through gauge-mediated RG mixing into Higgs-current operators. 
The additional hierarchy in perturbative order further strengthens the one from the experimental sensitivities, leading to substantially stronger constraints from $\tau$ decays than from $Z\to\tau\mu$ (cf.~again Fig.~\ref{fig:1D-constraints}). The operator $\mathcal{O}_{l e}$ is also constrained at one loop by $\tau\to\mu\gamma$, once more due to a finite contribution to the matrix element (see Appendix~\ref{app:observables}). Similar considerations apply to semileptonic operators involving light quarks, which can mediate, e.g., $\tau\to\mu\pi$, $\tau\to\mu\eta^{(\prime)}$, or $\tau\to\mu\rho$ at tree level.

In summary, $\tau$-lepton decays provide stronger constraints than $Z \to \tau \mu$ searches, both with current data and with the projected FCC-ee sensitivities. The results discussed above have been obtained under the simplifying assumption that only one operator at a time is active at the scale $\Lambda$. We also consider selected two-coefficient scenarios in Appendix~\ref{app:2D-SMEFT}, finding that this qualitative conclusion remains largely unchanged once correlations among Wilson coefficients are allowed. 


\subsection{LFV Higgs and $\tau$ decays}
\label{subsec:LFV-H-tau-decays}

\begin{figure*}[t]
    \centering
    \includegraphics[width=1\textwidth]{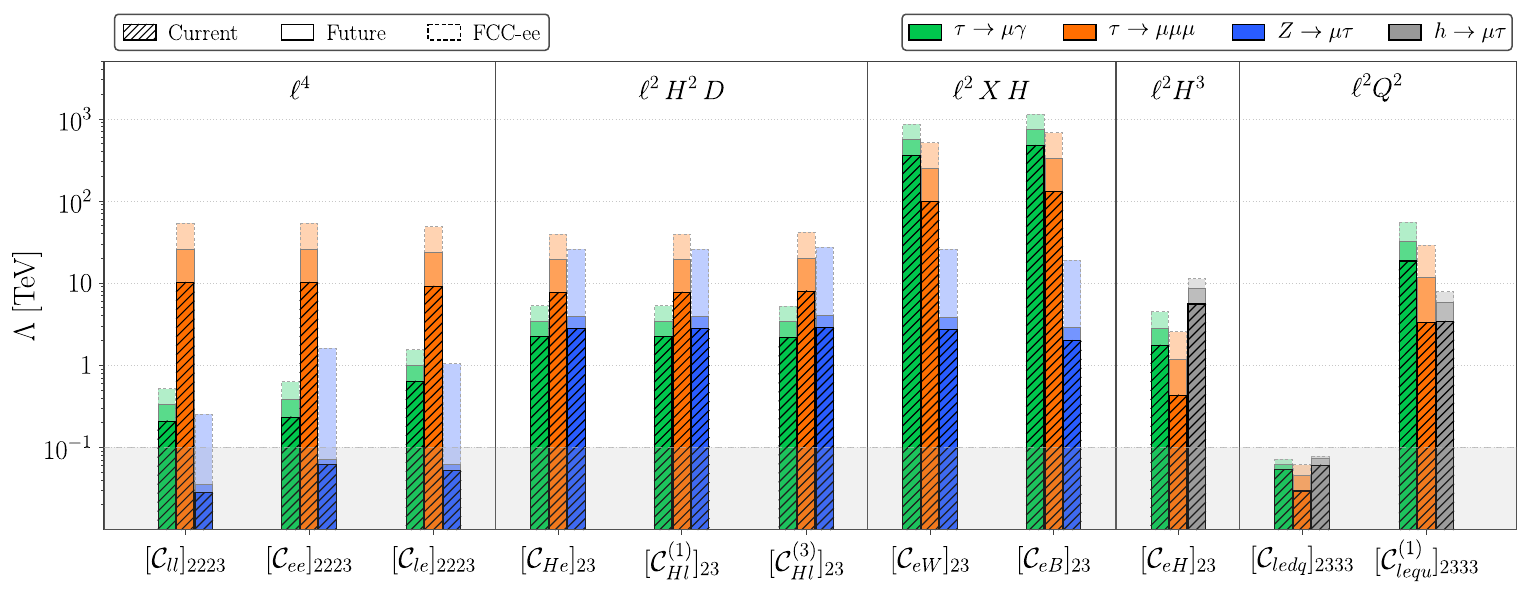}
    \caption{\small \sl Reach in the new-physics scale $\Lambda$ for $\mathcal{C}=1/\Lambda^2$ (where $\Lambda$ is also the scale from which the RG evolution is performed) at 95\% C.L.~for four-fermion, dipole, and Higgs-current operators in the $\tau \leftrightarrow \mu$ sector. Semileptonic operators are taken to have third-generation quark indices. Current bounds (hatched bars), and projections at Belle II/HL-LHC (solid bars) and FCC-ee (shaded bars) are shown. The shaded gray area corresponds to a new-physics reach below the electroweak scale, where the SMEFT is not the appropriate description.}
    \label{fig:1D-constraints}
\end{figure*}    

\begin{figure}[!t]
    \centering
    \begin{tikzpicture}[thick,>=stealth]
        \coordinate (c) at (0,1.2);
        \def\r{0.6}
        \draw[midarrow] (-2,0) node[left] {$\tau$} -- (-1,0);
        \draw[midarrow] (-1,0) -- (1,0);
        \node[above] at (0,0) {$\ell$};
        \draw[midarrow] (1,0) -- (2,0) node[right] {$\ell$};
        \draw[midarrow] (c) +(-30:\r) arc (-30:90:\r);
        \draw[midarrow] (c) +(90:\r) arc (90:210:\r);
        \draw[midarrow] (c) +(210:\r) arc (210:330:\r);
        \draw[vector] (c) + (-30:\r) -- (1,0);
        \draw[dashed] (c) + (210:\r) -- (-1,0);
        \draw[vector] (c) ++(90:\r) -- +(30:1) node[right] {$\gamma$};
        \draw[dotted] (-1,0) -- +(150:1) node[left] {$v$};
        \draw[dotted] (-1,0) -- +(120:1) node[above] {$v$};
        \fill (-1,0) circle (2pt);
        \node[below=3pt] at (-1,0) {$\mathcal{C}_{\substack{eH \\ \ell \tau}}$};
        \node[] at (1,0.6) {$\gamma$};
        \node[] at (-1,0.6) {$h$};
        \node[] at (0.8,1.3) {$t$};
    \end{tikzpicture}
    \caption{\small  \sl   Barr-Zee contribution to $\tau\to\ell\gamma$.}
    \label{fig:lfv-Higgs}
\end{figure}
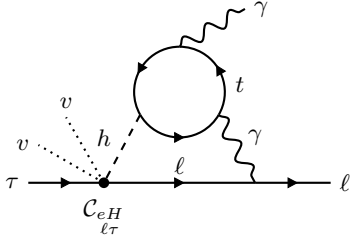

For Higgs-boson decays, we proceed in full analogy with the $Z$-boson decays discussed above. Among the operators in Table~\ref{tab:tau-ope-smeft}, $\mathcal{O}_{eH}$ is the only one that contributes to $h \to \tau \ell$ at tree level. The same operator contributes to $\tau\to\mu\gamma$ through loop effects that generate the electromagnetic dipole interaction, see e.g.~Ref.~\cite{Harnik:2012pb}. These include the well-known Barr-Zee contribution, illustrated in Fig.~\ref{fig:lfv-Higgs}, and two-loop RG mixing into the SMEFT dipole operators $\mathcal{O}_{eB}$ and $\mathcal{O}_{eW}$~\cite{Panico:2018hal,Allwicher:2025mmc}. Both contributions are discussed in detail in Appendix~\ref{app:observables}. 

The decay $\tau\to3\mu$ also receives contributions from $\mathcal{O}_{eH}$. The leading effects are induced by the same dipole interactions as $\tau\to \mu\gamma $ (through off-shell photon exchange), while the tree-level Higgs-exchange contribution is suppressed by the muon Yukawa coupling. Consequently, one expects $\mathcal{B}(\tau\to 3\mu)$ to be suppressed by more than two orders of magnitude with respect to $\mathcal{B}(\tau\to\mu\gamma)$. Indeed, we find that the Barr-Zee contributions lead to:
\begin{align}
\begin{aligned}    \dfrac{\mathcal{B}(\tau\to\mu\gamma)}{\mathcal{B}(h\to\tau \mu)}
   &  \simeq 1.1 \times 10^{-7}\,, \\[0.4em]
\dfrac{\mathcal{B}(\tau\to 3\mu)}{\mathcal{B}(h\to\tau \mu)}
   & \simeq 2.8 \times 10^{-10}\,.
 \end{aligned} 
\end{align} 

\noindent These scalings are reflected in the correlation plots in Fig.~\ref{fig:correlation-plots-H}, which show that searches for $h\to\tau\mu$ provide the strongest constraints on $\mathcal{C}_{eH}$ both with current data and with the projected future sensitivities. This is also apparent from the scale reach in Fig.~\ref{fig:1D-constraints}: at FCC-ee, $h\to\tau\mu$ probes scales of about $\Lambda = 11~\mathrm{TeV}$, compared with about $\Lambda = 4.5~\mathrm{TeV}$ from $\tau\to\mu\gamma$ decays and with $\Lambda =  2.6~\mathrm{TeV}$ from $\tau\to3\mu$ decays. 

The situation can change for operators that generate LFV Higgs interactions only at the loop level. To illustrate this possibility, we focus on the scalar semileptonic operators $\smash{\mathcal{O}_{lequ}^{(1)}}$ and $\mathcal{O}_{ledq}$ with third-generation quark flavor indices~\cite{Feruglio:2018fxo}. This choice maximizes the radiative contribution to $h \to \tau \mu$ through the large third-generation Yukawa couplings while avoiding tree-level contributions to semileptonic LFV $\tau$-lepton decays.

The contributions of these operators to $h \to \tau \mu$ decays arise through RG mixing into $\mathcal{C}_{eH}$~\cite{Jenkins:2013zja,Jenkins:2013wua}. These contributions are proportional to $\lambda\,y_q$ and $y_q^3$, where $q$ is the quark running in the loop, and $\lambda$ is the quartic Higgs coupling. For the $\smash{\mathcal{O}_{lequ}^{(1)}}$ operator, the leading contributions arise from the top-quark running in the loop ($\propto y_t^3$). Instead, for $\mathcal{O}_{ledq}$, the bottom quark runs in the loop, thus leading to a much smaller contribution ($\propto \lambda\, y_b$). 

The same operators contribute to LFV $\tau$-lepton decays through the radiative generation of the electromagnetic dipole. For $\mathcal{O}_{l equ}^{(1)}$, the leading effects involve one-loop RG mixing into the corresponding tensor operator, $\mathcal{O}_{l equ}^{(3)}$, followed either by further RG evolution into the SMEFT dipole operators or by one-loop matching into the LEFT dipole $O_{e\gamma}$. For $\mathcal{O}_{ledq}$, both genuine two-loop mixing into the SMEFT dipole operators as well as effectively three-loop contributions involving the chain $\mathcal{C}_{ledq} \to \mathcal{C}_{lequ}^{(1)} \to \mathcal{C}_{lequ}^{(3)} \to L_{e\gamma}$ are relevant. We include all these contributions in our numerical analysis. Further details are given in Appendix~\ref{app:observables}.

From Fig.~\ref{fig:1D-constraints}, we see that LFV $\tau$-lepton decays already provide the most stringent constraint on the semileptonic operator $\smash{\mathcal{O}_{lequ}^{(1)}}$. The FCC-ee projections translates into a substantial reach in the new-physics scale: at FCC-ee, $\tau\to\mu\gamma$ and $\tau\to3\mu$ probe scales of about $55~\mathrm{TeV}$ and $29~\mathrm{TeV}$, respectively, compared with about $8~\mathrm{TeV}$ from $h\to\tau\mu$. For $\mathcal{O}_{ledq}$, instead, the reach of all the observables considered here lies slightly below the electroweak scale, reflecting the poor sensitivity to this interaction. However, for this specific operator, useful constraints can be derived from the high-$p_T$ tails of $pp\to\mu\tau$ at the LHC~\cite{CMS:2022fsw}, which can probe operators in the TeV range with current data~\cite{Allwicher:2022gkm,Allwicher:2022mcg}.

\section{Conclusions}
\label{sec:outlook}

In this letter, we have investigated the prospects for probing LFV at future high-luminosity $e^+e^-$ facilities such as FCC-ee, assuming heavy new physics described within the SMEFT framework. Exploiting the unprecedented sample of approximately $2\times10^{11}$ $\tau$-lepton pairs expected from $Z$-boson decays, we have compared the sensitivity of $\tau$-lepton decays with that of searches for $Z$- and Higgs-boson decays. 
For LFV $Z$ interactions, we find that rare $\tau$ decays provide the stronger probe both with current data and with the projected FCC-ee sensitivities. In particular, $\tau\to\ell\gamma$ and $\tau\to3\ell$ can probe substantially higher effective new-physics scales than $Z\to \tau \ell$ searches and are also sensitive to a broader class of SMEFT interactions. 

The situation is more varied for LFV Higgs interactions. For new physics that modifies the SM Yukawa couplings, direct searches for $h\to\tau\ell$ provide the strongest sensitivity, since the corresponding effects in LFV $\tau$-lepton decays arise only at two-loop order. This hierarchy, however, is not generic. As an example, we have shown that for semileptonic operators that generate the LFV Higgs couplings only radiatively, $\tau$-lepton decays can become the leading probe.

Overall, our results strengthen the case for a dedicated LFV program in the $\tau$-lepton sector at FCC-ee. These processes can test new physics beyond the reach of LFV $Z$ and, in some cases, Higgs searches. These encouraging prospects motivate detailed Monte Carlo simulations to confirm the projected sensitivities, as well as studies of complementary probes, such as LFV $\tau$-lepton decays into hadrons, which we leave for future work.

\section*{Acknowledgments}

We thank Laura Zani for useful discussions on the Belle-II experimental inputs and Mogens Dam for clarifications on the FCC-ee projections in Ref.~\cite{Dam:2018rfz}. 
This project has received support from the Agence Nationale de la Recherche under the contract ANR-25-CE31-2504, from the European Union’s Horizon 2020 and Horizon Europe research and innovation programmes under the Marie Skłodowska-Curie grant agreements N$^\circ$~860881-HIDDeN and N$^\circ$~101086085-ASYMMETRY, respectively, from the IN2P3 (CNRS) Master Project HighPTflavor, from the USP-COFECUB project Uc Ph194-25/25, from the Generación de Conocimiento by MICIU/AEI/10.13039/501100011033 and FEDER (UE) under project PID2023-147706NB-I00, from the C\'esar Nombela Research Talent Attraction of the Community of Madrid under project 2023-T1/COM-29356, from the Deutsche Forschungsgemeinschaft under Germany’s Excellence Strategy EXC 2121 “Quantum Universe” – 390833306, as well as from the DFG grant 491245950.


\appendix

\section{Notation and conventions}
\label{app:conventions}

We work in the Warsaw basis for the SMEFT~\cite{Grzadkowski:2010es}, with quark and lepton doublets denoted by $q$ and $l$, respectively, and the SM weak singlets by $u,d$ and $e$. The covariant derivative is defined with the convention $D_\mu = \partial_\mu+i g_3 T^A A_\mu^A+i g_2 t^I W_\mu^I+i g_1 y B_\mu$, where $T^A$ are the $SU(3)_c$ generators, $t^I$ are the $SU(2)_L$ generators and $y$ is the hypercharge generator.

The Yukawa Lagrangian is defined as
\begin{align}
\mathcal{L}_{\mathrm{Yuk}}=-Y_u\,\bar{q} \widetilde{H} u-Y_d\,\bar{q}{H}d-Y_e\,\bar{l}{H}e + \mathrm{h.c.}\,, 
\end{align}
where $Y_{u}$ and $Y_d$ are the up- and down-quark Yukawa matrices, respectively, and $Y_{e}$ is the charged lepton Yukawa matrix. The sum over flavor indices is implicit. We work in the flavor basis where both the down-quark and charged-lepton Yukawa matrices are diagonal, while the up-quark Yukawa matrix reads
\begin{equation}
    Y_u = V^\dagger\,\mathrm{diag}(y_u,y_c,y_t)\,,
\end{equation}
where $V$ is the CKM matrix and $y_\psi = \sqrt{2} m_\psi / v$ are the Yukawa couplings at the electroweak scale, $m_\psi$ denotes the corresponding quark mass, $v=(\sqrt{2}G_F)^{-1/2}$ is the electroweak vacuum expectation value and $G_F$ stands for the Fermi constant.
We write the Higgs potential as
\begin{align}
    V(H) = \lambda \left(H^\dagger H - \frac{v^2}{2}\right)^2 \,,
\end{align}
such that $m_h =v \sqrt{2\lambda}$.


\section{Expressions for observables}
\label{app:observables}

In this Appendix, we collect the expressions for the branching fractions of LFV $Z$- and Higgs-boson decays, as well as $\tau$-lepton decays, used in our numerical analysis. We also provide the RG and matching contributions relevant to the SMEFT operators considered in the main text.

\paragraph*{$Z$ decays.} The $Z\to \ell_i \ell_j$ decays can be described, after the electroweak symmetry breaking, by
\begin{align}
    \mathcal{L}_Z &\supset Z_\mu \, \dfrac{m_Z}{v} \, g_{V_L}^{ij}\, \big{(} \bar{\ell}_{Li}\gamma^\mu \ell_{Lj}\big{)} \\
    &+ Z_{\mu\nu}\, \dfrac{g_{D_L}^{ij}}{\sqrt{2}\,v} \, \bar{\ell}_{Li} \sigma^{\mu\nu} \ell_{Rj} + (L\leftrightarrow R) +\mathrm{h.c.}\,, \nonumber
\end{align}

\noindent where flavor indices are fixed. This Lagrangian leads to the decay rate~\cite{Brignole:2004ah,Plakias:2023esq},~\footnote{Notice that our expression for the dipole contributions to the decay rate differs from Refs.~\cite{Crivellin:2013hpa,Calibbi:2021pyh}.}
\begin{align}
    \Gamma(Z\to \ell_i^- \ell_j^+)= &\frac{m_Z^3}{24\pi v^2}\bigg[|g_{V_L}^{ij}|^2+|g_{D_L}^{ij}|^2 + (L\leftrightarrow R)\bigg]\,, 
\end{align}
where lepton masses are neglected, and the effective couplings are given at tree level by
\begin{align}
    g_{V_L}^{ij} &={v^2}\,\mathcal{C}_{\substack{Hl\\ ij}}^{(1+3)}\,, \qquad\qquad
    g_{V_R}^{ij} =  {v^2} \,\mathcal{C}_{\substack{He\\ ij}} \,,  \\*[0.4em]
    g_{D_L}^{ij} &= \big{(}g_{D_R}^{ji}\big{)}^\ast = - {v^2}\,(s_W\, \mathcal{C}_{\substack{eB\\ij}}+c_W \, \mathcal{C}_{\substack{eW\\ij}})\,, \nonumber
\end{align}

\noindent where $s_W=\sin \theta_W$ and $c_W=\cos\theta_W$, and the effective coefficients are evaluated at the electroweak scale.

Four-lepton operators do not contribute to $Z\to \ell_i \ell_j$ decays at tree level, but they induce LFV $Z$ couplings via RG mixing into the Higgs-current operators~\cite{Feruglio:2016gvd,Feruglio:2017rjo,Cornella:2018tfd}. For simplicity, we use the following notation,
\begin{align}
    \dot{\mathcal{C}} \equiv \dfrac{\mathrm{d}\mathcal{C}}{\mathrm{d}\log\mu}\,.
\end{align}

\noindent At one loop, the relevant terms are, in the convention where flavor indices are symmetrized~\cite{Alonso:2013hga},
\begin{align}
         \dot{\mathcal{C}}_{\substack{Hl \\ ij}}^{(1)} &= -\frac{g_1^2}{48\pi^2} \Big( 2 \,\mathcal{C}_{\substack{l e\\ ijkk}} + 2 \,\mathcal{C}_{\substack{l l\\ ijkk}}+ \mathcal{C}_{\substack{l l\\ ikkj}}\\*
         &\qquad\qquad\qquad\quad+\mathcal{C}_{\substack{l l\\ kjik}} +2 \,\mathcal{C}_{\substack{l l\\ kkij}}\Big)+\cdots\,,\nonumber\\[0.35em]
        \dot{\mathcal{C}}_{\substack{Hl \\ ij}}^{(3)} &=+ \frac{g_2^2}{48\pi^2} \Big( \mathcal{C}_{\substack{l l\\ ikkj}}+ \mathcal{C}_{\substack{l l\\ kjik}}\Big)+\cdots\,,\\[0.35em]
         \dot{\mathcal{C}}_{\substack{He \\ ij}} &= -\frac{g_1^2}{24\pi^2} \Big( \mathcal{C}_{\substack{ee\\ ijkk}} + \mathcal{C}_{\substack{ee\\ ikkj}}+ \mathcal{C}_{\substack{ee\\ kjik}}\nonumber\\
         &\qquad\qquad\qquad\quad+\mathcal{C}_{\substack{ee\\ kkij}} + \mathcal{C}_{\substack{l e\\ kkij}}\Big)+\cdots\,,
\end{align}
where the ellipses denote contributions that are not relevant for the scenarios considered here, cf.~Table~\ref{tab:tau-ope-smeft}. For the four-lepton operator contribution to LFV $Z$ decays, we retain only the logarithmically enhanced contribution generated by this RG mixing.

\paragraph*{Higgs decays.} New physics contributions to the Higgs-boson coupling to leptons can be described in the SMEFT by the operator $\mathcal{O}_{eH}$ defined in Table~\ref{tab:tau-ope-smeft}. After electroweak symmetry breaking, this operator generates an effective Yukawa coupling
\begin{equation}
   \mathcal{L}_H \supset -{y^{ij}_{e}} \,h\,\bar{\ell}_{Li} \ell_{Rj}+\mathrm{h.c.}\,,
\end{equation}

\noindent in addition to a direct contribution to the charged-lepton mass matrix, denoted by $M_e$. More specifically, at the electroweak scale~\cite{Harnik:2012pb},
\begin{align}
    \sqrt{2} \, y_e &= Y_e  - \dfrac{3}{2}\mathcal{C}_{eH}\,v^2\,, \\[0.35em]
    \sqrt{2} \, \dfrac{M_e}{v} &= Y_e - \dfrac{1}{2}{\mathcal{C}_{eH}\,v^2} \,, \nonumber
\end{align}

\noindent where flavor indices are omitted. Therefore, the charged-lepton mass matrix and the Yukawa couplings depend on different combinations of the bare SM Yukawa $Y_e$ and the contributions from dimension-six operators $\mathcal{C}_{eH}$, which are not necessarily simultaneously diagonalizable. The Higgs couplings accessible experimentally can then be expressed in terms of the fermion masses as follows,
\begin{align}
    \label{eq:eff-yuk}
    y_{e} =  \dfrac{M_e}{v}- \mathcal{C}_{\substack{eH}}\,\dfrac{v^2}{\sqrt{2}} \,,
\end{align}

\noindent where $M_e$ must be diagonalized to give fermion masses. The LFV Higgs decay rate is given by
\begin{equation}
    \Gamma(h\to \ell_i^- \ell_j^+) = \dfrac{m_h}{16\pi} \bigg{[} |y^{ij}_e|^2 + |y^{ji}_e|^2 \bigg{]}\,,
\end{equation}

\noindent where we recall that the above expressions hold at $\mu \simeq \mu_{\mathrm{ew}}$ scale.

The semileptonic operators $\mathcal{O}_{ledq}$ and $\smash{\mathcal{O}_{l equ}^{(1)}}$ contribute to $h \to \ell_i \ell_j$ via RG evolution from $\Lambda$ down to the electroweak scale~\cite{Feruglio:2018fxo}. The relevant one-loop RG equations read~\cite{Jenkins:2013zja,Jenkins:2013wua},~\footnote{We note that the bare Yukawa $Y_e$ also receives RG contributions from these semileptonic operators, but these contributions are reabsorbed in Eq.~\eqref{eq:eff-yuk}.}
\begin{align}
 \dot{\mathcal{C}}_{\substack{eH\\ ij}} &=  \frac{N_c}{4\pi^2} \left[\lambda \big{(}Y_d\big{)}_{33}-\big{(}Y_d Y_d^\dagger Y_d\big{)}_{33}\right]  \mathcal{C}_{\substack{l e d q\\ ij33}}  \\*
&- \frac{N_c}{4\pi^2} \left[\lambda\big{(}Y_u^\dagger\big{)}_{33}-\big{(}Y_u^\dagger Y_u  Y_u^\dagger\big{)}_{33}\right] \mathcal{C}_{\substack{l e q u\\ ij33}}^{(1)} + \dots\,,  \nonumber
\end{align}

\noindent where $N_c$ denotes the number of colors, and we have kept only the contributions from third-generation quarks, as they generate the leading effects. Notice that the term proportional to three insertions of the Yukawas is dominant for $\smash{\mathcal{C}_{lequ}^{(1)}}$, i.e.~$Y_u^\dagger Y_u Y_u^\dagger$, whereas the one proportional to $\lambda \, Y_d$ gives the leading effect for $\mathcal{C}_{ledq}$ due to the smallness of $y_b$.

\paragraph*{Radiative $\tau$ decays.} 

The purely leptonic LFV $\tau$ decays considered here can be generally expressed in terms of the LEFT Lagrangian,
%
\begin{align}   
    \mathcal{L}_{\mathrm{LEFT}} \supset \sum_a {L}_a\,{O}_a\,,
\end{align}
%

\noindent where ${O}_a$ are the effective operators invariant under $SU(3)_c\times U(1)_{\mathrm{em}}$ and $L_a$ are the relevant effective coefficients. In the scenario that we consider, the relevant low-energy operators are the four-lepton ones and the dipole operators~\cite{Kuno:1999jp},
\begin{align}
O_{\substack{ee \\ ijkl}}^{\rm VXY}&=(\overline{e}_{Xi}\gamma^\mu e_{X j})\,(\overline{e}_{Y k}\gamma_\mu e_{Y l})\,,\\[0.35em] O_{\substack{e\gamma \\ij}}&=\overline{e}_{Li}\,\sigma^{\mu\nu} e_{Rj} F_{\mu\nu}\,, \nonumber
\end{align}
where $X,Y = L,R$ denote the chirality of the charged leptons and $F_{\mu\nu} = \partial_\mu A_\nu - \partial_\nu A_\mu$ is the electromagnetic field-strength tensor. The corresponding LEFT Wilson coefficients are denoted by $L_{ee}^\mathrm{VXY}$ and $L_{e\gamma}$. 

The decay rate for $\tau\to\mu\gamma$ can be written in our scenario as follows,
\begin{align}
    \Gamma(\tau\to \mu \gamma)= &\frac{(m_\tau^2 - m_\mu^2)^3}{4\pi \,m_\tau^3}\Bigg\{\Big|L_{\substack{ e\gamma \\ 23}} - \frac{e}{16\pi^2}\sum_i m_i \,L_{\substack{ ee \\ 2ii3}}^{\rm VLR}\Big|^2 \nonumber\\
    &\quad + \Big|L_{\substack{ e\gamma \\ 32}}^* - \frac{e}{16\pi^2}\sum_i m_i \, L_{\substack{ ee \\ i32i}}^{\rm VLR}\Big|^2\Bigg\}\,,
    \label{eq:taumugamma}
\end{align}
where all coefficients are understood to be at $\mu = m_{\tau}$.
The terms proportional to the four-lepton Wilson coefficient $L_{\substack{ ee}}^{\rm VLR}$ arise from the finite one-loop matrix element of the corresponding operator. 
At the electroweak scale, the LEFT Wilson coefficients entering Eq.~\eqref{eq:taumugamma} are related to the SMEFT Wilson coefficients defined in Table~\ref{tab:tau-ope-smeft} by~\cite{Dekens:2019ept},
\begin{align}
    L_{\substack{ ee \\ 2ii3}}^{\rm VLR} &= \mathcal{C}_{\substack{ l e\\ 2ii3}} + \delta_{i2} \frac{v^2}{4 m_Z^2}(g_1^2 - g_2^2) \,\mathcal{C}_{\substack{ H e\\ 23}}\, \label{eq:loop-matching-VLR} \\
    &\quad\quad+\,\delta_{i3} \frac{g_1^2\,v^2}{2 m_Z^2} \,\mathcal{C}_{\substack{ H l \\ 23}}^{(1+3)}\,,\nonumber\\[0.45em]
    L_{\substack{ ee \\ i32i}}^{\rm VLR} &=  \mathcal{C}_{\substack{ l e\\ i32i}} + \delta_{i3} \frac{v^2}{4 m_Z^2}(g_1^2 - g_2^2) \,\mathcal{C}_{\substack{ H e\\ 23}}\,\label{eq:loop-matching-VLR-2} \\
    &\quad\quad+\,\delta_{i2} \frac{g_1^2\,v^2}{2 m_Z^2} \,\mathcal{C}_{\substack{ H l \\ 23}}^{(1+3)}\,,\nonumber\\[0.45em]
    L_{\substack{ e\gamma \\ ij}}&= \frac{v}{\sqrt{2} }\Big(c_W \,\mathcal{C}_{\substack{eB\\ij}} - s_W \,\mathcal{C}_{\substack{ eW \\ ij}}\Big)\label{eq:loop-matching-egamma}\\*
    &+\frac{e}{48\pi^2}\Bigg[ 4 m_i\, s_W^2 \, \mathcal{C}_{\substack{ H e\\ ij}} + 2 m_j\, (s_W^2-c_W^2) \, \mathcal{C}_{\substack{ H l\\ ij}}^{(1)} \nonumber\\*
    &+ m_j (3c_W^2+7 s_W^2) \, \mathcal{C}_{\substack{ H l\\ ij}}^{(3)} - 24 m_t \log\Big(\frac{m_Z^2}{m_t^2}\Big)\,\mathcal{C}_{\substack{l e q u\\ ij33}}^{(3)}\Bigg]\,\nonumber\\*
    &+ \cdots,\nonumber 
\end{align}
where, for each SMEFT operator considered, we display the 
leading contribution up to one-loop order.  However, for specific operators, higher-order contributions can be phenomenologically relevant~\cite{Panico:2018hal,Plakias:2023esq}. In particular, this is the case for $\mathcal{O}_{eH}$, $\mathcal{O}_{ledq}$ and $\smash{\mathcal{O}_{lequ}^{(1)}}$:

\begin{itemize}[leftmargin=*]
    \item The leading contribution from the operator $\mathcal{O}_{eH}$ to $\tau \to \mu \gamma$ appears at two loops. We include both the finite Barr-Zee contribution and the two-loop RG mixing into the SMEFT dipole operators. The former gives, at the Higgs mass scale~\cite{Allwicher:2025mmc}
\begin{align}
    L_{\substack{ e\gamma \\ ij}} (m_h)=\frac{1}{(16\pi^2)^2} \frac{8\sqrt{2}}{3} e^3 v \, F_{\rm loop}\Big(\frac{m_t^2}{m_h^2}\Big) \mathcal{C}_{\substack{ e H\\ ij}}(m_h)\,,
\end{align}
where
\begin{align}
    F_{\rm loop}(x) = \frac{x(1-2x)}{w}&\bigg[{\rm Li}_2\Big(1-\frac{1-w}{2x}\Big)\\*
    & -{\rm Li}_2\Big(1-\frac{1+w}{2x}\Big)\bigg] + x(2+\log x)\,,\nonumber
\end{align}
with $w\equiv\sqrt{1-4x}$. 
In addition, $\mathcal{O}_{eH}$ contributes through two-loop running into the dipole~\cite{Panico:2018hal,Allwicher:2025mmc},
\begin{align}
    \begin{pmatrix}
        \dot{\mathcal{C}}_{eB} \\ \dot{\mathcal{C}}_{eW}
    \end{pmatrix}_{ij}&= \frac{3g_2^3}{8(16\pi^2)^2}\begin{pmatrix}
        t_W - 3 t_W^3 \\
        1-t_W^2
    \end{pmatrix}\mathcal{C}_{\substack{ eH \\ ij}}+\dots\,,
\end{align}
where $t_W=\tan \theta_W$.  Both contributions are included in our analysis. 

\item For $\smash{\mathcal{O}_{lequ}^{(1)}}$, the effect in $\tau \to \mu \gamma$ is due to one-loop mixing into the tensor operator $\mathcal{O}_{lequ}^{(3)}$~\cite{Plakias:2023esq},  
\begin{align}
  \dot{\mathcal{C}}_{\substack{l equ \\ 2333}}^{(3)} &= \frac{1}{16\pi^2}\left[\frac{5g_1^2+3g_2^2}{8}\right] \mathcal{C}_{\substack{l equ \\ 2333}}^{(1)}+\dots
\label{eq:lequ1_into_lequ3}
\end{align}

\noindent Once generated, $\mathcal{O}_{lequ}^{(3)}$ contributes to the electromagnetic dipole in two ways. First, it mixes into the SMEFT dipoles via 
\begin{align}
    \dot{\mathcal{C}}_{\substack{eB \\ 23}} &= \frac{5}{8\pi^2} g_1 (Y_u)_{33}^* \, \mathcal{C}_{\substack{l equ \\ 2333}}^{(3)}+\cdots\,,\\
    \dot{\mathcal{C}}_{\substack{eW \\ 23}} &= -\frac{3}{8\pi^2} g_2 (Y_u)_{33}^* \, \mathcal{C}_{\substack{l equ \\ 2333}}^{(3)}+\cdots\,.
\end{align}
Second, it gives a finite one-loop contribution to the SMEFT to LEFT matching onto $L_{e \gamma}$~\cite{Dekens:2019ept}. 

\item 
For $\mathcal{O}_{ledq}$, several contributions of different perturbative orders compete. First, $\mathcal{O}_{ledq}$ mixes into the SMEFT dipoles via two-loop RG equations~\cite{EliasMiro:2021jgu},
\begin{align}
   \begin{pmatrix}
        \dot{\mathcal{C}}_{eB} \\ \dot{\mathcal{C}}_{eW}
    \end{pmatrix}_{ij}&= \frac{y_b \,g_2^3\, N_c}{8(16\pi^2)^2}\begin{pmatrix}
        t_W - 5 \,t_W^3/3 \\
        1-t_W^2
    \end{pmatrix}\mathcal{C}_{\substack{ l edq \\ ij33}}\,. 
\end{align}
In addition, it mixes at one loop into $\mathcal{O}_{lequ}^{(1)}$,
\begin{align}
    \dot{\mathcal{C}}_{\substack{l equ \\ 2333}}^{(1)} &= -\frac{1}{4\pi^2}(Y_d)_{33}(Y_u)_{33} \, \mathcal{C}_{\substack{l edq \\ 2333}}+\dots\,,
\end{align}
and therefore inherits the mechanisms described above for $\mathcal{O}_{lequ}^{(1)}$. Although formally of higher order than the direct two-loop mixing into the SMEFT dipoles, these contributions can be numerically relevant for the scale separation at hand, and are therefore included in our analysis~\cite{Plakias:2023esq}.    
\end{itemize}

\vspace{15pt}
\begin{table*}[!t]
    \centering
    \renewcommand{\arraystretch}{2.2}
    \setlength{\tabcolsep}{3.5pt}
    \begin{tabular}{ll|ccc|ccc|cc|ccc}
        \hline\hline
        & &
        \multicolumn{3}{c|}{Four-lepton} &
        \multicolumn{3}{c|}{Higgs-current} &
        \multicolumn{2}{c|}{Dipole} &
        \multicolumn{3}{c}{Higgs and semileptonic}
        \\
        Process & Dataset
        & $\mathcal{C}_{ll}$
        & $\mathcal{C}_{ee}$
        & $\mathcal{C}_{l e}$
        & $\mathcal{C}_{He}$
        & $\mathcal{C}_{Hl}^{(1)}$
        & $\mathcal{C}_{Hl}^{(3)}$
        & $\mathcal{C}_{eW}$
        & $\mathcal{C}_{eB}$
        & $\mathcal{C}_{eH}$
        & $\mathcal{C}_{l edq}$
        & $\mathcal{C}_{l equ}^{(1)}$
        \\
        \hline\hline

        \multirow{3}{*}{$\tau\to\mu\gamma$}
        & Current
        & $0.2$ & $0.2$
        & $0.6$
        & $2.2$
        & $2.2$
        & $2.2$
        & ${357}$
        & ${473}$
        & $1.8$
        & \belowEW
        & $19$
        \\
        & Belle II
        & $0.3$ & $0.4$
        & $1.0$
        & $3.4$
        & $3.4$
        & $3.4$
        & ${558}$
        & ${739}$
        & $2.8$
        & \belowEW
        & $33$
        \\
        & FCC-ee
        & $0.5$ & $0.6$
        & $1.6$
        & $5.3$
        & $5.3$
        & $5.2$
        & ${860}$
        & ${1138}$
        & $4.5$
        & \belowEW
        & $55$
        \\
        \hline

        \multirow{3}{*}{$\tau\to3\mu$}
        & Current
        & ${10}$
        & ${10}$
        & ${9.1}$
        & ${7.7}$
        & ${7.7}$
        & ${8.0}$
        & $98$
        & $130$
        & $0.4$
        & \belowEW
        & $3.3$
        \\
        & Belle II
        & $26$
        & $26$
        & $23$
        & $19$
        & $19$
        & $20$
        & $247$
        & $328$
        & $1.2$
        & \belowEW
        & $12$
        \\
        & FCC-ee
        & $53$
        & $54$
        & $49$
        & $39$
        & $40$
        & $42$
        & $510$
        & $677$
        & $2.6$
        & \belowEW
        & $29$
        \\
        \hline

        \multirow{3}{*}{$Z\to\mu\tau$}
        & Current
        & \belowEW
        & \belowEW
        & \belowEW
        & $2.8$
        & $2.8$
        & $2.9$
        & $2.7$
        & $2.0$
        & -- & -- & --
        \\
        & HL-LHC
        & \belowEW
        & \belowEW
        & \belowEW
        & $4.0$
        & $4.0$
        & $4.1$
        & $3.9$
        & $2.9$
        & -- & -- & --
        \\
        & FCC-ee
        & $0.3$
        & $1.6$
        & $1.0$
        & $26$
        & $26$
        & $27$
        & $26$
        & $19$
        & -- & -- & --
        \\
        \hline

        \multirow{3}{*}{$h\to\mu\tau$}
        & Current
        & -- & -- & -- & -- & -- & -- & -- & --
        & $5.6$
        & \belowEW
        & $3.4$
        \\
        & HL-LHC
        & -- & -- & -- & -- & -- & -- & -- & --
        & ${8.7}$
        & \belowEW
        & $5.8$
        \\
        & FCC-ee
        & -- & -- & -- & -- & -- & -- & -- & --
        & ${11}$
        & \belowEW
        & $7.9$
        \\
        \hline\hline
    \end{tabular}%

    \caption{\small \sl Lower bounds on the new-physics scale
    $\Lambda$ in TeV at~95$\%$ C.L., obtained by setting the indicated Wilson
    coefficient to $1/\Lambda^2$ at the scale $\Lambda$ and all other coefficients to zero.
    Current constraints are compared with projections for
    Belle~II or the HL-LHC, as appropriate for each process,
    and FCC-ee. A dash indicates that the corresponding process does not constrain the indicated Wilson coefficient at the order considered. Entries marked with \belowEW~correspond to an inferred reach below $\mu_{\mathrm{ew}}=100~\mathrm{GeV}$, where the SMEFT interpretation is not applicable. }
 \label{tab:numerical_scale_bounds}
\end{table*}

\paragraph*{$\tau \to 3\mu$.} The decay rate of $\tau \to 3\mu$ reads 
\begin{align}
    \Gamma(\tau&\to 3\mu)= \frac{m_\tau^5}{1536\pi ^3}\Bigg\{2\Bigg(\Big|L_{\substack{ ee \\ 2223}}^{\rm VLL}\Big|^2 + \Big|L_{\substack{ ee \\ 2223}}^{\rm VRR}\Big|^2\Bigg)\nonumber\\*
    &+\Big|L_{\substack{ ee \\ 2223}}^{\rm VLR}\Big|^2+\Big|L_{\substack{ ee \\ 2322}}^{\rm VLR}\Big|^2 \nonumber\\*
    &+\frac{64 e^2}{m_\tau^2} \Bigg[\log\Big(\frac{m_\tau}{m_\mu}\Big) - \frac{11}{8}\Bigg]\Big(|L_{\substack{ e\gamma \\ 23}}|^2 + |L_{\substack{ e\gamma \\ 32}}|^2 \Big)  \nonumber\\&+\frac{8 e}{m_\tau}\Bigg({\rm Re}\left[L_{\substack{ e\gamma \\ 23}} \,\Big(2 L_{\substack{ ee \\ 2223}}^{\rm VLL} +L_{\substack{ ee \\ 2322}}^{\rm VLR}\Big)^*\,\right] \nonumber\\*
    &+ {\rm Re}\Big[L_{\substack{ e\gamma \\ 32}}^{\ast} \,\Big(2 L_{\substack{ ee \\ 2223}}^{\rm VRR} +L_{\substack{ ee \\ 2223}}^{\rm VLR}\Big)^*\,\Big]\Bigg) \Bigg\}\,, 
\end{align}
where we used the convention in which redundant operator structures are removed, see e.g.~Ref.~\cite{Fuentes-Martin:2020zaz}. 
For $L_{ee}^{\rm VLR}$ and $L_{e\gamma}$, the relevant matching relations are given in Eqs.~\eqref{eq:loop-matching-VLR}, \eqref{eq:loop-matching-VLR-2} and \eqref{eq:loop-matching-egamma}. For the remaining LEFT Wilson coefficients, the tree-level matching to the SMEFT yields
\begin{align}
    L_{\substack{ ee \\ 2223}}^{\rm VLL} &=  \frac{v^2}{4 m_Z^2}(g_1^2 - g_2^2) \,\mathcal{C}_{\substack{ H l \\ 23}}^{(1+3)} + \mathcal{C}_{\substack{ ll \\ 2223}}\,,\\
    L_{\substack{ ee \\ 2223}}^{\rm VRR
    } &=  \frac{g_1^2\,v^2}{2 m_Z^2} \, \mathcal{C}_{\substack{ H e \\ 23}} + \mathcal{C}_{\substack{ ee \\ 2223}}\,\,.
\end{align}

\section{Two-dimensional constraints}
\label{app:2D-SMEFT}

\begin{figure*}[!h]
    \centering
\includegraphics[width=1.\textwidth]
{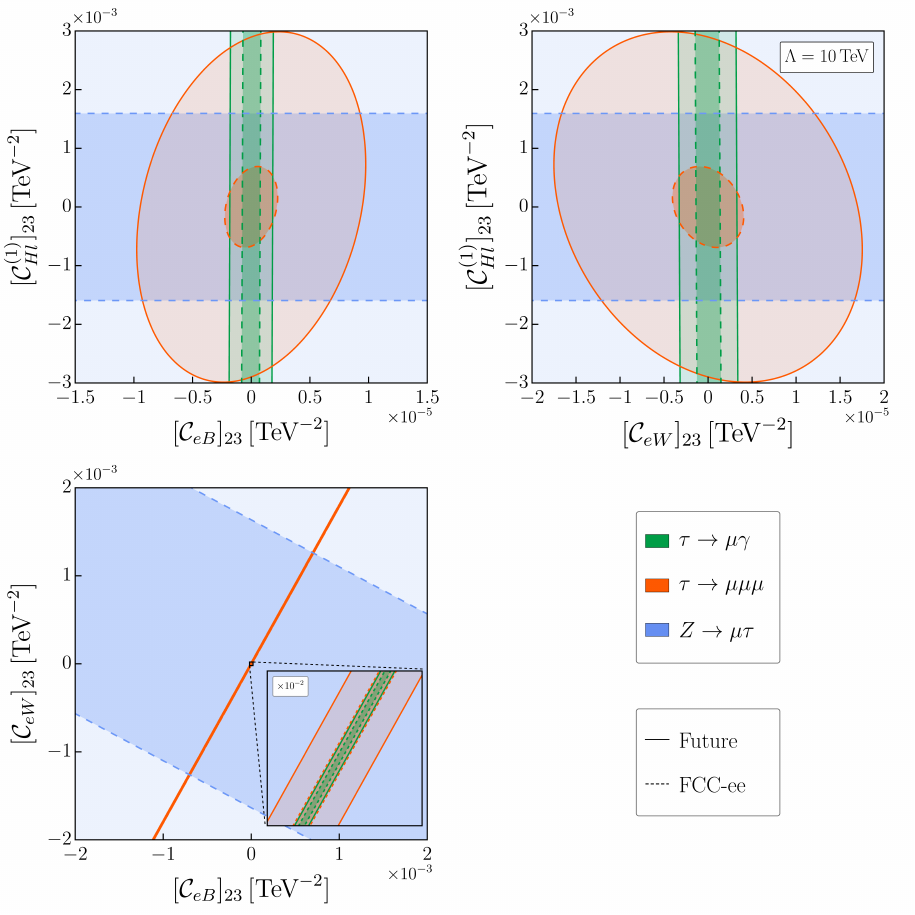}
    \caption{\small \sl Two-dimensional constraints on the $\lbrace \mathcal{C}_{Hl}^{(1)},\,\mathcal{C}_{eB},\,\mathcal{C}_{eW} \rbrace$ Wilson coefficients, with flavor indices $23$, which directly enter $Z\to\mu\tau$ decays at tree level. Projections at Belle II/HL-LHC (solid lines) and FCC-ee (dashed lines) are shown. For illustration, the ultraviolet scale $\Lambda$ from which running effects are taken into account is fixed to be $10$~TeV.}
    \label{fig:2Dgrid}
\end{figure*}

The results shown in the main text are all obtained under the assumption of a single SMEFT coefficient being active at a time at the ultraviolet scale $\Lambda$.
In Fig.~\ref{fig:2Dgrid}, we show the constraints for different Wilson coefficient pairs, for which potential flat directions may appear.
As one can see, both for dipoles and Higgs-current operators, the possible flat directions are lifted by combining $Z$-boson and $\tau$-lepton decays.
This is because the latter always receive a large photon-dipole contribution.
The allowed sizes of the Wilson coefficient, however, can vary greatly, as highlighted by the zoomed-in region, and coherently with Fig.~\ref{fig:1D-constraints}.
A very similar result can be obtained in the case of Higgs decays, in the plane $\cC_{lequ}^{(1)}$--$\cC_{ledq}$, where the $\tau$ decays probe an independent direction.
In general, operators with the same field content but different chirality structures (such as the four-lepton operators) do not interfere, and hence no flat direction can appear in the fit.
In the case of $\cC_{Hl}^{(1,3)}$, the combination orthogonal to the $Z$ coupling is probed by the one-loop contributions in radiative $\tau$ decays.

\bibliographystyle{JHEP}
\bibliography{refs}

\end{document}